# Batch-Fabricated Cantilever Probes with Electrical Shielding for Nanoscale Dielectric and Conductivity Imaging

Yongliang Yang[1,2], Keji Lai[2], Qiaochu Tang[1], Worasom Kundhikanjana[2], Michael A. Kelly[2], Kun Zhang[1], Zhi-xun Shen[2#] and Xinxin Li[1*]

[1]State Key Laboratory of Transducer Technology, Shanghai Institute of Microsystem and Information Technology, Chinese Academy of Sciences, Shanghai 200050, People's Republic of China
[2] Department of Applied Physics and Geballe Laboratory for Advanced Materials, Stanford University, Stanford, California 94305, USA

[*] Author to whom any correspondence should be addressed (for the formation of the probe). Email: xxli@mail.sim.ac.cn
[#] Author to whom any correspondence should be addressed (for the probe surface imaging). Email: zxshen@stanford.edu

**Abstract**

This paper presents the design and fabrication of batch-processed cantilever probes with electrical shielding for scanning microwave impedance microscopy. The diameter of the tip apex, which defines the electrical resolution, is less than 50 nm. The width of the stripline and the thicknesses of the insulation dielectrics are optimized for a small series resistance (< 5 $\Omega$) and a small background capacitance (~ 1 pF), both critical for high sensitivity imaging on various samples. The coaxial shielding ensures that only the probe tip interacts with the sample. The structure of the cantilever is designed to be symmetric to balance the stresses and thermal expansions of different layers so that the cantilever remains straight under variable temperatures. Such shielded cantilever probes produced in the wafer scale will facilitate enormous applications on nanoscale dielectric and conductivity imaging.

# 1. Introduction

Near-field scanning microwave microscopy has been demonstrated to study the microscopic dielectric and conductivity properties of materials using a sub-wavelength probe tip [1-5]. As the tip scans over the sample surface, variation of the local electrical properties results in changes of the tip-sample impedance, which are then detected by electronics to form microwave impedance microscopy (MIM) images with a spatial resolution comparable to the diameter of the tip apex [6-12]. Due to its direct access to important local electrodynamic properties such as the complex dielectric permittivity and permeability [4], this technique has been utilized to study both fundamental electron physics, such as electronic phase transitions [11, 13] and quantum Hall effect [14], and applied science, such as the electrical properties of biological samples [15, 16] at the microscopic level.

The widespread applications of microwave microscopy, however, have been largely hindered by the premature system design. Early implementations of microwave microscopes were configured as a sharp needle tip protruding from a cavity or transmission line resonator [17-20]. Despite the relatively high sensitivity [3, 19, 21-23], such systems usually require special bulky scanners and the tip apex easily becomes blunted because of the lack of feedback control [3, 19]. Micro-fabricated cantilever probes on atomic-force microscope (AFM) platforms provide an elegant solution to the above difficulties [6-9, 12, 24-28]. Thanks to the advanced MEMS technology, cantilevers with sub-100 nm tip apex are routinely achieved [29-32] and well preserved by the good tip-sample distance control [26]. Different from conventional AFM tips, cantilevers probes for microwave imaging are much more difficult to be fabricated due to the following reasons. First, the tip apex should be sharp for high spatial resolution. The dimension of the tip apex defines the spatial extension of quasi-static electric field, which sets the MIM spatial resolution [6]. The well-established anisotropic etching of crystalline silicon to form a sharp tip [29, 30] cannot be applied in MIM tips because the MIM tips usually are low resistivity metal such as platinum and gold, which cannot be anisotropically etched in the same way. Second, a low-resistance metallic path from the bonding pad (for signal input / output) to the tip apex is needed to minimize the loss. The background capacitance between this signal line and the ground should also be kept small for better electrical sensitivity because MIM measures the sample introduced impedance variation upon the base impedance [12]. Third, the signal line has to be electrically shielded in order to reduce the stray fields and noise pickup. In other words, the center conductor should be surrounded by dielectrics and shield metals. Finally, such a sandwich structure with stacking metal/dielectric layers is highly susceptible to the bimorph behavior [33]. It is imperative to balance the stresses and thermal expansions of different layers so that the cantilever is straight and remains so at variable temperatures.

The complexity of microwave probes presents a practical challenge for MEMS fabrications. In our previous work, we developed a silicon nitride based cantilever structure with aluminum center path and shields [6, 12]. After the micro-fabrication, a platinum tip was deposited onto the cantilever free end by focused ion beam (FIB). While considerable success has been made, the spatial resolution of 100~200 nm is limited by the relatively big FIB tip. More importantly, this one-at-a-time FIB deposition is not scalable for batch process, which continues as the major obstacle in advancing the MIM technique. In this paper, we demonstrate the fabrication process of cantilever probes with electrical shielding for MIM applications. The gold/titanium tips are shown to be sharper (< 50 nm in diameter) than the FIB tips. The width of the stripline and the thickness of the dielectrics are optimized for small series resistance (< 5 Ω) and small background capacitance (~ 1 pF). The electrical shielding ensures that only the tip interacts with the sample. The layer structure of the cantilever body is symmetric to prevent bending when the temperature is varied. No post-fabrication process is needed so the probes can be uniformly produced in the wafer scale.

## 2. Fabrication process

A schematic of the designed microwave probe is shown in figure 1. The main body of the cantilever is made of plasma enhanced chemical vapor deposited (PECVD) $Si_3N_4$. The TiW/Au metal tip on the free end of the cantilever is connected to the wire bond pad by a conducting path buried inside two $Si_3N_4$ layers. Both the front (tip side) and back sides of the cantilever are covered by shield metals, which are electrically grounded in the microwave measurements.

The detailed fabrication processes are shown in figure 2. The starting materials are double-side polished (100) silicon wafers. Pyramidal pits are etched in aqueous KOH etchant to form the tip molds, where four (111) surfaces meet at the apex [figure 2(a)]. The pit is further sharpened by a low temperature wet oxidation at 950 °C for 8 hours. The measured oxide thickness is about 1 μm. Due to the compressive stress in the silicon oxide, the thickness of the oxide at the apex is less than that in a flat surface [34], which further sharpens the apex [figure 2(b)]. The first metal layer, which consists of 50 nm of TiW, 400 nm of Au, and another 50 nm of TiW, is deposited in DC mode of Denton Discovery Sputtering System. After lithography, the metals are patterned by wet etching to form the tip metal (inside the pit), the front shield metal, and the 100 μm × 200 μm wire bond pad. The TiW is etched in $H_2O_2$ (30%, 50 °C) and the Au is etched in gold etchant (5% $I_2$ +10% KI +85% $H_2O$, ~20 °C) [figure 2(c)]. A dielectric layer of 0.8 μm PECVD $Si_3N_4$ is then deposited and patterned as the cantilever body. We put another layer of 1 μm PECVD $SiO_2$ on the die to further increase the dielectric thickness here for reducing the background capacitance. The dielectric layers on the pit, the bond pad and the via-holes (located near the cantilever end) are removed for electrical connection [figure 2(d)]. Next, the second metal layer (50 nm TiW / 800 nm Au / 50 nm TiW) is patterned into the center conducting path, 6 μm-width on the cantilever and 14 μm-width on the die, to connect the metal tip and the bond pad [figure 2(e)]. The wafer is then covered by a second 0.8 μm PECVD $Si_3N_4$ layer [figure 2(f)] and the backside shield metal [same metal stacks as front metal, figure 2(g)], both patterned into desirable shapes. Note that via-holes going through the dielectrics are made to electrically short the front and back shields, which are shown in the top views of figures 2(d)-(g). At the same time, a handle wafer with thermal oxides on both sides is fabricated with KOH etched trenches. 3um Bisbenzocyclobutene (BCB,

CYCLOTENE 3022-46 from The Dow Chemical Company) is spin coated on the handle wafer and baked on a hotplate at 105 °C for 2 minutes to stabilize the film. Then the handle wafer and the device wafer are aligned to have the trenches on top of the cantilever and bonded in Karl Suss SB-6. In $N_2$ ambient, the temperature rises to 250 °C in 30 minutes and then stays at 250 °C for 60 minutes, finally cools to room temperature in 60mins. The device wafer and the handle wafer are glued together by BCB [figure 2(h)]. Such bonded wafers are anisotropically etched in aqueous tetramethyl ammonium hydroxide (TMAH) to completely remove the silicon in the device wafer, as well as the silicon trenches (not protected by $SiO_2$) in the handle wafer [figure 2(i)]. Then the back side of the cantilever is covered by spray-coating 3 μm photoresist, when the wafers are mounted on a home-made holder to support the probes without damaging the cantilever. After etching the oxide on the device wafer by buffered oxide etchant, the photoresist is removed in acetone and the cantilever probes are finally released [figure 2(j)]. Figure 2(k) shows the front-view (tip side) of the probe and the cross-sectional views of cantilever (A-A') and tip (B-B'). The cantilever structure is symmetric about the center plane except for the small center conductor. In our process, the handle wafer brings a handle on the back side of the cantilever so that the probe can be mounted on the z-scanner of MIM and the tip on the front side can easily land on the sample surface. We emphasize that the entire fabrication process is suitable for batch-production.

As shown in figure 3(a), hundreds of probes are fabricated on a 4 inch wafer. The dimensions are is 3.4 mm × 1.6 mm for the dies and the 300 μm × 50 μm for the cantilevers. The scanning electron micrographs (SEM) in figures 3(b) and 3(c) show the bond pad on the die, the cantilever, and the pyramidal metal tip. A close-up view of the tip in figure 3(d) shows a sharp apex with a diameter less than 50 nm.

**3. Testing results**

Great care was taken in the design and fabrication of the microwave probes to minimize both the series resistance ($R_s$) of the center conductor and its capacitance ($C_{tip}$) to ground. Small $R_s$ is desired to reduce the loss in the signal line and increase the sensitivity. In our case, the thick center conducting path and optimized conducting path width keep the measured $R_s$ below 5Ω, which is much smaller than the doped Si trace in other implementations [24, 25, 32, 35]. Since the tip-sample interaction is essentially a tiny modulation to the tip capacitance, we have used sufficiently thick dielectrics and optimized conducting path width to minimize $C_{tip}$ to ~ 1pF without compromising the mechanical properties. The low $R_s$, $C_{tip}$, as well as the shielded structure (which will be discussed later), are critical for the exquisite MIM results described below.

Figure 4 shows the schematic of MIM setup. 1GHz microwave signals are delivered to the metal probe tip and the reflected microwave signals contain the local dielectric and conductivity information of sample material. The microwave electronics detect the imaginary and real components of the tip-sample impedance and output as MIM-Im and MIM-Re signals. The surface topography is simultaneously obtained by the AFM laser feedback. The characterization and analysis of the MIM system are detailed in References [9, 12] and not repeated here. Standard samples are scanned with our batch fabricated probes. All the images are obtained with normal AFM settings. The scanning velocity is 20 μm/s and the contact force between the tip and sample is 1 nN.

Figure 5 demonstrates the ability to perform conductivity imaging by the new MIM probe. The selectively doped Si sample here is similar to the one used in Reference [13] except that the substrate is nearly intrinsic (ρ > 1000 Ω·cm). As illustrated in figure 5(a), the heavy implantation of phosphorus ions results in slight surface damage in the implanted regions. The minor surface roughness, although discernible in the AFM image in figure 5(b), is totally overwhelmed by the strong conductivity contrast between the implanted and un-implanted areas in the MIM-Im [figure 5(c)] and MIM-Re [figure 5(d)] images. The data quality is comparable to that taken by the FIB tips. Analysis of the microwave signals is detailed in Reference [13] and not repeated here.

The ability to perform sub-surface dielectric imaging by the new probes is shown in figure 6. The same polished $Al_2O_3$ / $SiO_2$ sample in Reference [6] was used for this purpose. The flat sample surface after the polishing enables the demonstration of unambiguous dielectric contrast between the 120 nm sub-surface $Al_2O_3$ and $SiO_2$ layers. As expected, the microwave contrast is purely in the imaginary part [figure 6(c)], with only noise in the MIM-Re channel [figure 6(d)].

The shielded cantilever structure is very important for the local electrical imaging. For comparison, we show the MIM images of an etched silicon sample taken by an unshielded commercial conductive AFM tip, which is a metal coated Si probe, and our shielded tip. Square patterns were etched on a high resistivity silicon wafer [figure 7(a) and (b)], producing a sample with only topographic variation and no electrical difference between the squares and the substrate. For the unshielded conductive AFM tip, the entire cantilever probe interacts with the sample. As the tip moves up and down to follow the surface profile, the distance between the cantilever and the sample changes accordingly. Therefore, a large topography-induced contrast is seen in the capacitive MIM-Im channel [figure 7(c)]. We note that the exposed conducting path also picks up enormous noise from the environment, as shown clearly in figure 7(c). For our shielded probe, on the other hand, only the pyramidal probe tip interacts with the sample. The topographic artifact is thus much reduced, showing essentially no contrast between the squares and the substrate except at the step edges [figure 7(d)]. In other words, we have proved that electrical shielding is critical to minimize topographic contributions in the final MIM signals. The noise level in figure 7(d) is also much lower than that in figure 7(c), again reflecting the significance of shielding.

With a spring constant of about 1 N/m and sub-50 nm tip diameter, our probes show excellent topography performance, which is the same as most commercial AFM contact tips and much better than FIB tips. A sample with arrays of Ni nano-dots, 100nm in diameter and 50nm in height, has been used to test the sharpness of the tip. As seen in figure 8(b), the AFM image taken by our tip clearly shows superior topographic resolution compared with the blurred image obtained by the previous FIB tip [figure 8(c)].

The higher electrical spatial resolution of our tip is verified by imaging an exfoliated graphene piece on the standard $SiO_2$ / Si substrate. For comparison, AFM and MIM-Im images taken by the batch-fabricated tip and the FIB tip are shown in figures 9(a)-(d). Due to the already very high conductivity (~ $10^5$ S/m), the step edge between single- and multi-layer graphene is not obvious in the MIM images. Since the topographic contribution on the single-layer graphene side is minimal, one could extract the electrical spatial resolution from the signal rising edge, as shown in figure 9(e). It is clear that the rising of MIM signal taken by the new TiW/Au tip is much steeper than the FIB Pt tip, consistent with the sharper image in figure 9(b). In figure 9(e), a spatial resolution of ~80 nm, comparable with the tip diameter, is extracted from the rising edge across the boundary of two distinct materials.

The multi-layer structure of our cantilevers introduces another complication in the mechanical and thermal properties. If not designed properly, the different internal stresses and thermal expansions from different layers may result in severe bending either right after the release or under elevated / cryogenic temperatures. Since the metals and $Si_3N_4$ would inevitably have different stresses and thermal expansion coefficients, the key here is to employ a symmetric design about the center plane for self-compensation of these effects. Note that the narrow center conductor only covers 1/10 of the cantilever and has negligible effects on the mechanical and thermal properties. The cantilever is symmetric about the center plane [as shown in figure 10(a)]. The top half and bottom half of the cantilever have the same internal stresses and thermal expansions. Both stress-induced and temperature-induced bending moments are well balanced. Thus our cantilever is straight at both room temperature [figure 10 (b)] and 400 K [figure 10 (c)]. These probes also show satisfactory performance at low temperatures for condensed matter physics research, which will be discussed elsewhere.

## 4. Conclusions

Cantilever probes with electrical shielding for scanning microwave impedance microscopy have been design and batch fabricated. With KOH etching and low temperature oxidation processes, ultra-sharp metal tip with apex diameter less than 50 nm has been realized on the cantilever. The width of the stripline and the thickness of the insulation dielectrics are optimized for small series resistance and background capacitance. The shielding metals are integrated on both sides of the cantilever. The fabricated probes show excellent performances in both AFM and MIM imaging. The symmetric layer structure ensures straight cantilevers for variable temperature experiments. Such wafer-scale production of shielded cantilever probes finally removes the obstacle for widespread nanoscale dielectric and conductivity imaging applications.

## 5. Acknowledgments

The authors from Shanghai Institute of Microsystem and Information Technology appreciate the project support from Chinese 973 Program (2011CB309503) and NSFC Project (91023046, 61021064). For the authors from Stanford University, the work is supported by NSF grants DMR-0906027 and Center of Probing the Nanoscale PHY-0425897. Xinxin Li also thanks Korean WCU project (R32-2009-000-20087-0). We would like to thank Nahid Harjee, Alexandre Haemmerli, Beth Pruitt and David Goldhaber-Gordon for their valuable suggestions on probe fabrication and helpful discussions.

Captions:

Figure 1. (a) 3D schematic of the design shielded probe. (b) Front (tip-side) view of the cantilever, showing the position of the metal tip. (c) Back side of the cantilever, showing the back shield and the buried center conducting path.

Figure 2. Process flow: (a) KOH etching. (b) Thermal oxidation to sharpen the pit. (c) Deposition and patterning of the front metal. (d) Process of the first PECVD $Si_3N_4$ layer. (e) Fabrication of the center path. (f) Deposition of the second PECVD $Si_3N_4$. (g) Deposition and patterning of the back shield metal. (h) Wafer bonding. (i) TMAH etching to remove the silicon. (j) Release of the cantilever. In (a) – (j), the left panels are cross-sectional views and the right panels are top views. (k) Front-view (tip side) of the probe and cross-sectional views of the cantilever (A-A') and tip apex (B-B').

Figure 3. Pictures of the completed cantilever probes. (a) Finished 4" wafer with hundreds of probes. (b) SEM image of the handle chip. (c) Close-up view of the cantilever and metal tip. (d) Side view of the tip and its sharp apex. The diameter of the tip apex is less than 50 nm.

Figure 4. Schematic of MIM setup. The microwave electronics detect the imaginary and real components of the tip-sample impedance and output as MIM-Im and MIM-Re signals. Surface topography is simultaneously obtained by the AFM laser feedback.

Figure 5. (a) Schematic of the doped Si sample. (b) AFM, (c) MIM-Im, and (d) MIM-Re images of the same sample. The heavy implantation results in slight surface damage here. Higher MIM-Im (brighter) signals are seen in the implanted, thus more conducting areas than the un-implanted regions, while the latter show higher loss (MIM-Re) signals.

Figure 6. (a) Schematic of the polished $Al_2O_3$ / $SiO_2$ sample. (b) AFM, (c) MIM-Im, and (d) MIM-Re images of the same sample. Clear sub-surface dielectric contrast is observed in the MIM-Im channel.

Figure 7. (a) Schematic of the etched Si sample. (b) AFM, (c) MIM-Im image obtained by a commercial conductive AFM tip without shielding. Topography profile induces larger signal in the MIM-Im image. (d) MIM-Im image obtained by our well-shielded tip, showing little contrast between the squares and the substrate. The MIM-Re images (not shown) from both tips show no contrast because of the high resistivity silicon sample.

Figure 8. AFM performance. (a) SEM image of the nano-dots. (b) Clear AFM image taken by our tip. (c) Blurred AFM image taken by the FIB tip.

Figure 9. (a) AFM, and (b) MIM images taken by our batch-fabricated probe. (c) AFM, and (d) MIM images taken by the previous FIB probe. (e) Line profiles in (b) and (d), showing the better spatial resolution of the current tips.

Figure 10. (a) Schematic of cantilever side view. Side views of the cantilever at (b) room temperature (293 K) and (c) 400 K.

**Figure 1**

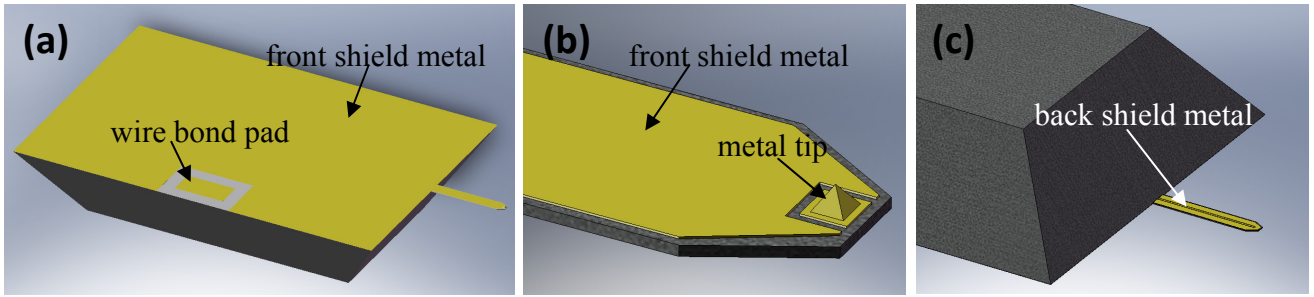

**Figure 2**

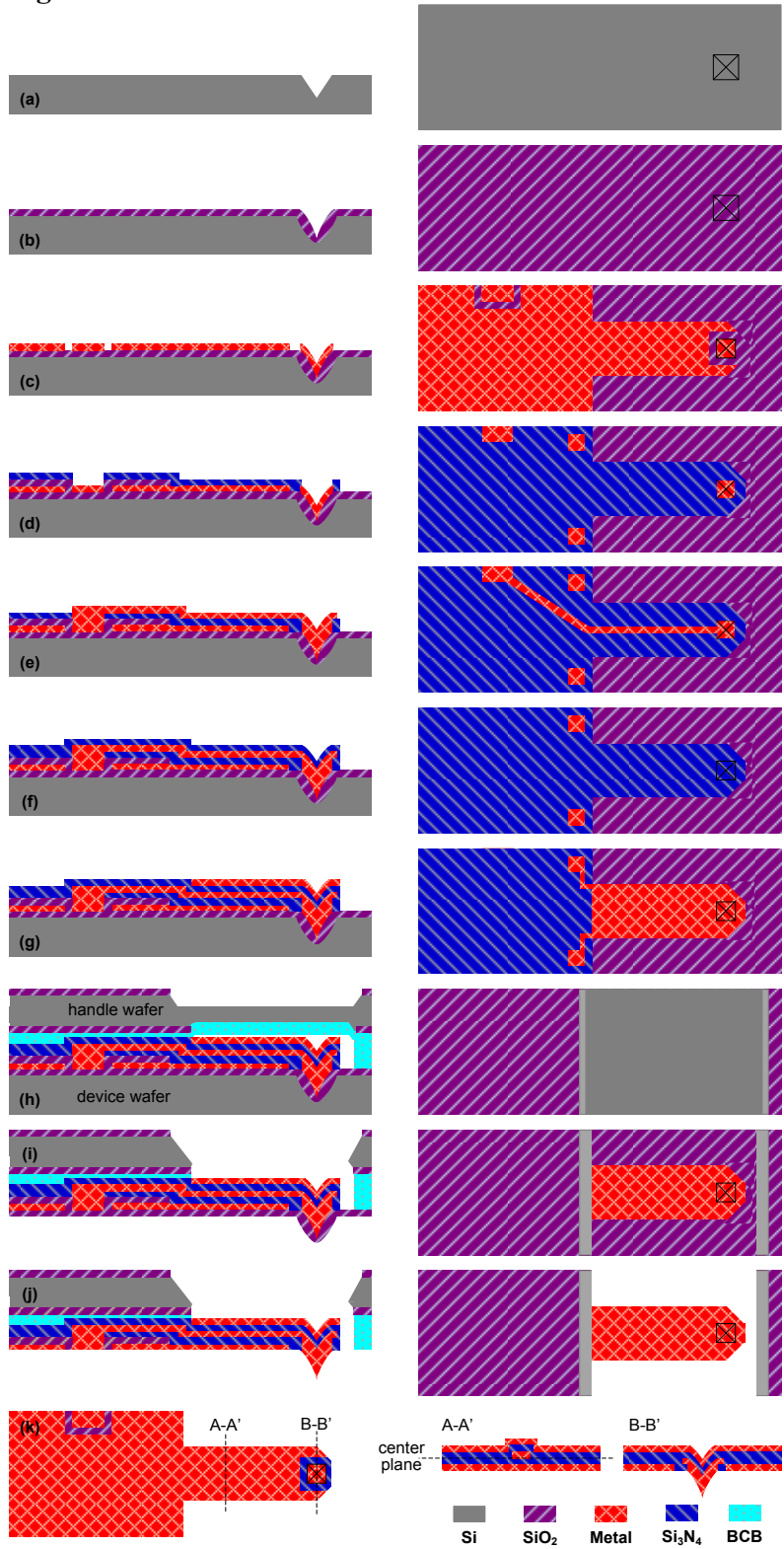

**Figure 3**

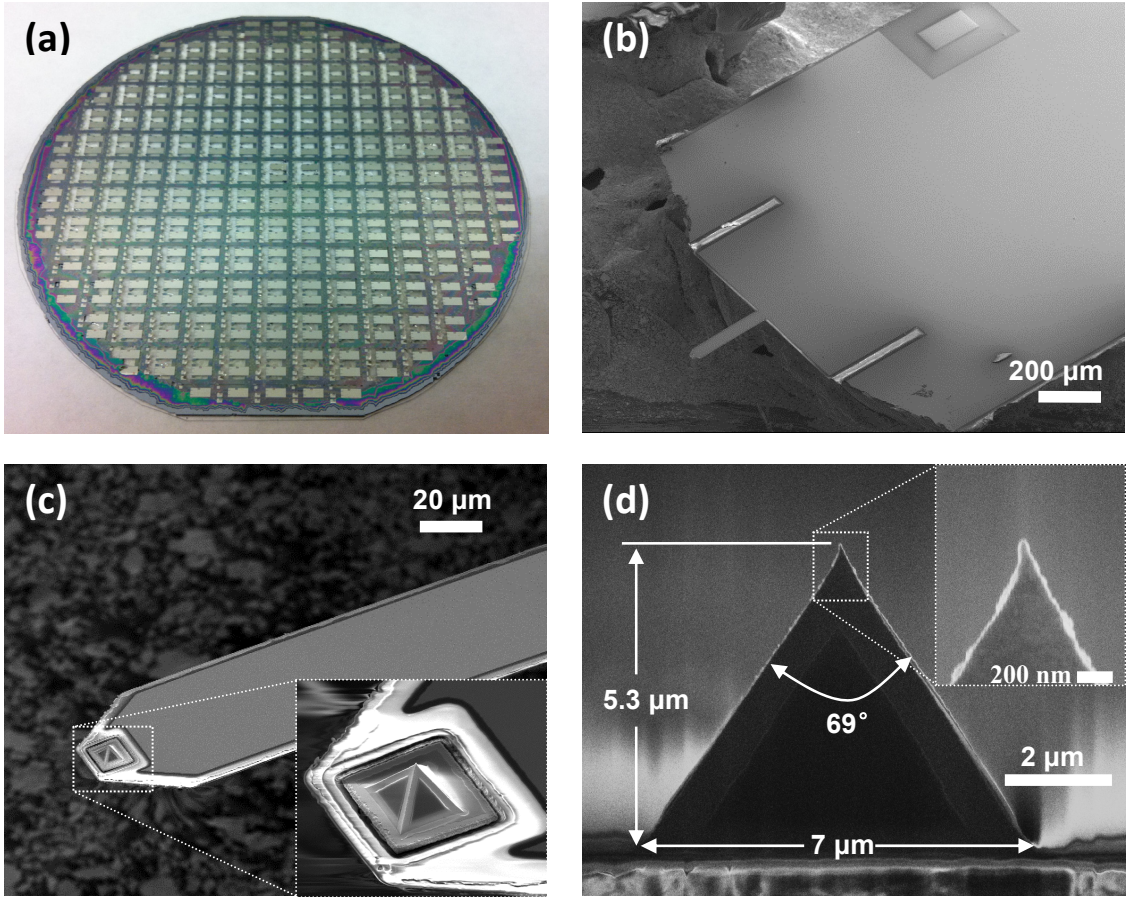

**Figure 4**

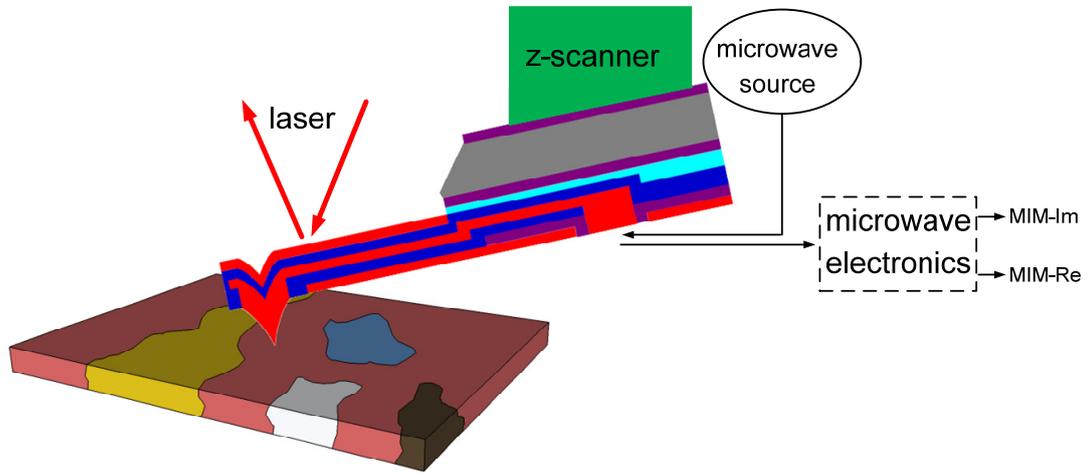

**Figure 5**

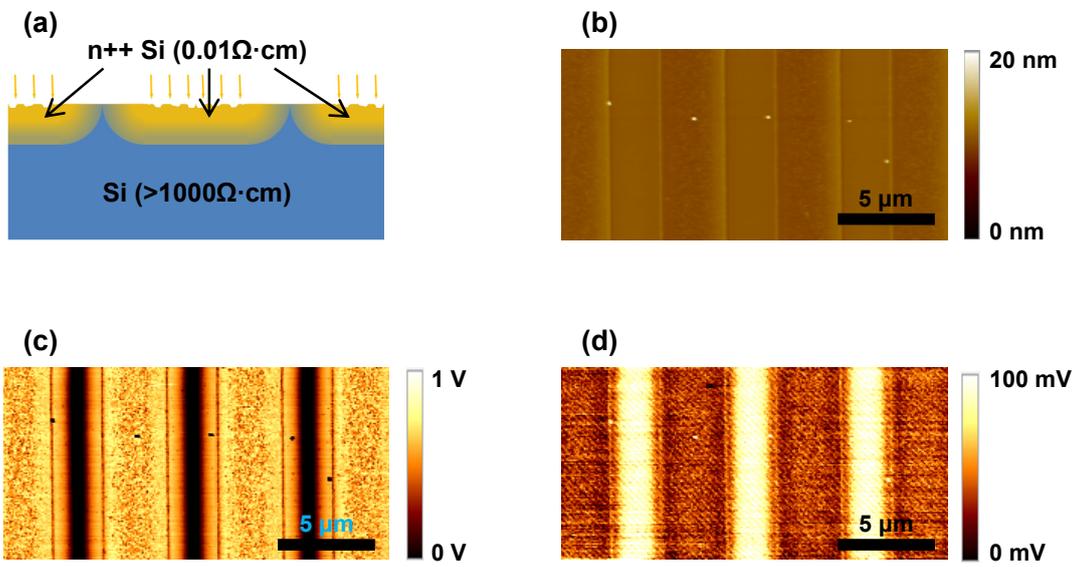

**Figure 6**

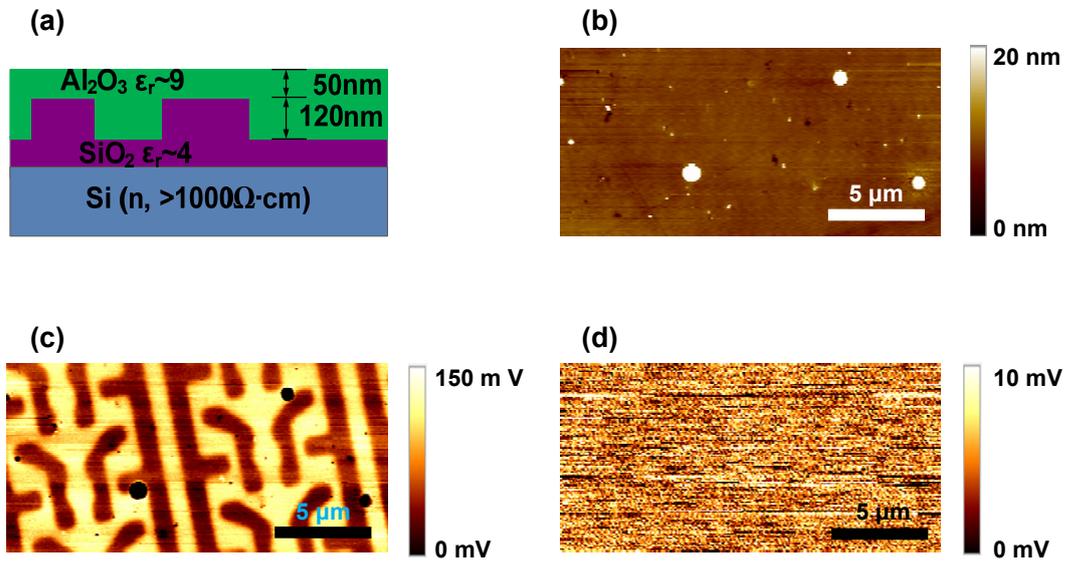

**Figure 7**

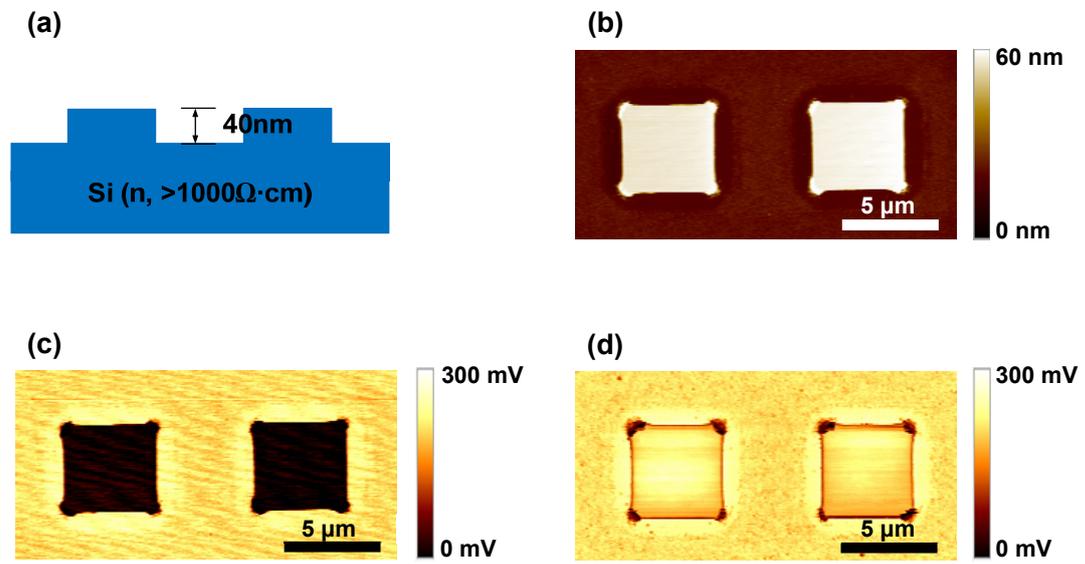

**Figure 8**

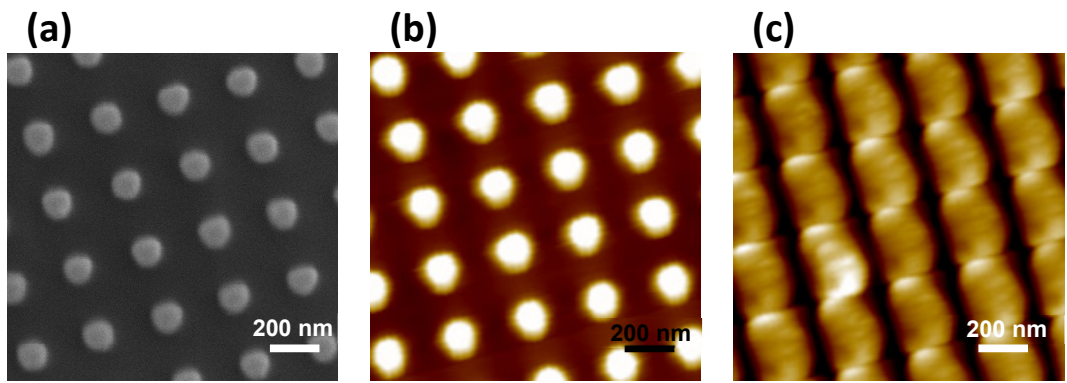

**Figure 9**

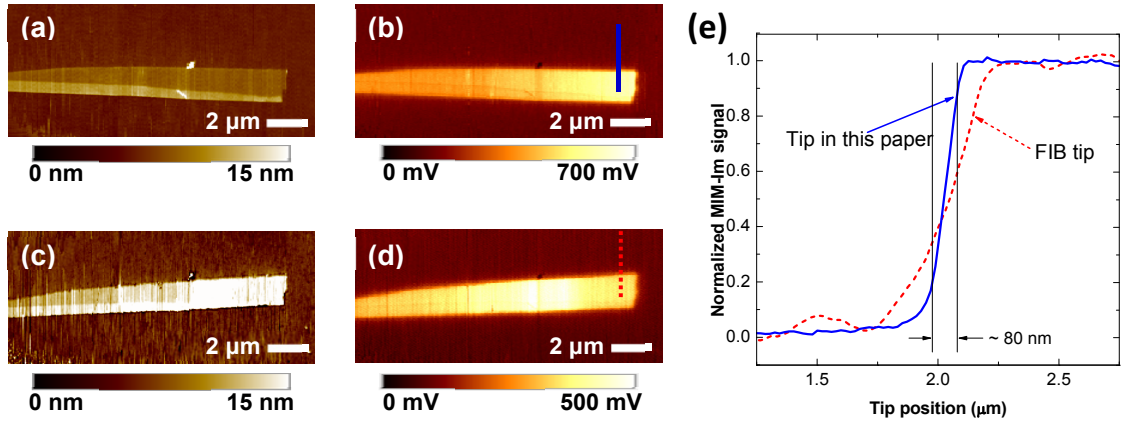

**Figure 10**

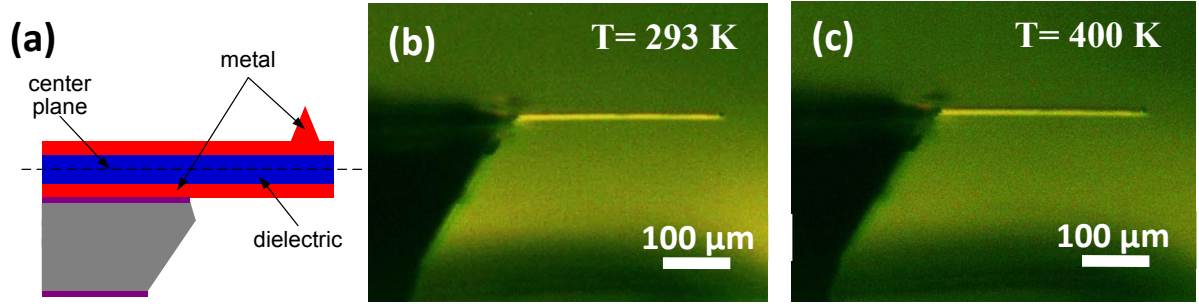